\documentclass[a4paper]{article}
\usepackage[left=2.5cm, right=2.5cm, top=2.5cm, bottom=2.5cm]{geometry}

\newcommand{\changefontsize}{\fontsize{12}{16}\selectfont}

\changefontsize

\usepackage{cmbright}
\usepackage[T1]{fontenc}

\usepackage[utf8]{inputenc}
\usepackage{amssymb}
\usepackage{amsfonts}
\usepackage{amsmath}
\usepackage{fontsmpl}
\usepackage{graphicx}
\usepackage{epsfig}
\usepackage[utf8]{inputenc}
\usepackage[english]{babel}
 
\usepackage{amsthm}
\usepackage{subcaption}
\usepackage{epsfig}

\usepackage[table,xcdraw]{xcolor}  
 \usepackage{booktabs} 
\usepackage{caption} 
\usepackage{subcaption}
\usepackage{rotating}

\usepackage[subfigure]{tocloft}

\usepackage{tikz}
\usepackage{subcaption}
\usepackage{xcolor}

\usepackage{booktabs}
\usepackage{multirow}
\usepackage{rotating}
\usepackage{forest}
\usepackage{xcolor}

\usepackage{graphicx}
\usepackage{booktabs}
\usepackage{float}

\usepackage{amsmath}

\title{The Future of Work and Capital: Analyzing AGI in a CES Production Model}
\author{Pascal Stiefenhofer\\ Department of Economics, Newcastle University\\pascal.stiefenhofer@newcastle.ac.uk}
\date{January 2025}

\begin{document}

	\maketitle

\begin{abstract}
The integration of Artificial General Intelligence (AGI) into economic production represents a transformative shift with profound implications for labor markets, income distribution, and technological growth. This study extends the Constant Elasticity of Substitution (CES) production function to incorporate AGI-driven labor and capital alongside traditional inputs, providing a comprehensive framework for analyzing AGI's economic impact. 

Four key models emerge from this framework. First, we examine the substitution and complementarity between AGI labor and human labor, identifying conditions under which AGI augments or displaces human workers. Second, we analyze how AGI capital accumulation influences wage structures and income distribution, highlighting potential disruptions to labor-based earnings. Third, we explore long-run equilibrium dynamics, demonstrating how an economy dominated by AGI capital may lead to the collapse of human wages and necessitate redistributive mechanisms. Finally, we assess the impact of AGI on total factor productivity, showing that technological growth depends on whether AGI serves as a complement to or a substitute for human labor.

Our findings underscore the urgent need for policy interventions to ensure economic stability and equitable wealth distribution in an AGI-driven economy. Without appropriate regulatory measures, rising inequality and weakened aggregate demand could lead to economic stagnation despite technological advancements. Moreover this research suggests a renegoation of the Social Contract.
\end{abstract}

\


\section{Introduction}
Artificial General Intelligence (AGI) is an advanced artificail intelgigence (AI) system capable of performing intellectual tasks with adaptability, autonomous reasoning, and recursive self-improvement. Unlike narrow AI, AGI generalizes across domains, applies abstract knowledge, and optimizes its own algorithms\cite{morris2024}. Once realized, AGI will function as both an autonomous economic agent and an accumulative capital asset, necessitating a re-evaluation of traditional production models\cite{agrawal2019},\cite{brynjolfsson2023}.
	
AGI’s integration into economic production disrupts conventional models reliant on human labor and physical capital. To capture its dual role, we extend the Constant Elasticity of Substitution (CES) production function \cite{arrow1961,zarembka1970,yankovyi2021} to incorporate human labor (\(L_h\)), AGI labor (\(L_{AGI}\)), traditional capital (\(K\)), and AGI capital (\(K_{AGI}\)). This enables a rigorous analysis of AGI’s effects on labor market dynamics, capital accumulation, and long-run equilibrium.
	
AGI labor (\(L_{AGI}\)) is infinitely scalable, incurs near-zero marginal costs, and challenges wage-based economic structures. It can either complement human labor, enhancing productivity, or substitute it entirely, exacerbating labor displacement. Network effects further amplify its economic impact, necessitating revised macroeconomic models to address employment shifts and income redistribution.
	
AGI capital (\(K_{AGI}\)) consists of physical infrastructure (e.g., AI hardware) and intangible algorithmic assets. Unlike traditional capital, AGI capital self-improves, potentially yielding super-linear returns and defying the principle of diminishing marginal productivity. Its dual role as both capital and labor blurs conventional economic distinctions, requiring an expanded production framework.
	
The adoption of AGI will profoundly reshape economic structures, influencing wages, productivity, and wealth distribution. Its effect on total factor productivity hinges on whether AGI complements or displaces human labor—either driving sustained economic growth or causing stagnation due to weakened demand. Managing these disruptions requires proactive policy interventions to ensure the equitable distribution of AGI-driven economic gains.

The current Social Contract\cite{rousseau1762} is iunsuffuciently equipped to address the profound societal disruptions that  AGI is poised to bring. As AGI advances, society must renegotiate its foundational agreements to reflect emerging economic and social conditions. This research is essential because AGI has the potential to redefine labor markets, capital allocation, and overall productivity.

A critical question remains: Will AGI augment human labor or replace it? The answer will determine long-term economic trajectories—whether AGI drives expansion or exacerbates inequalities. By identifying these dynamics, policymakers can craft informed strategies to mitigate risks, ensure economic stability, and equitably distribute the benefits of AGI. A proactive approach will be crucial in shaping a more inclusive and resilient economy, ensuring that AGI serves as a force for progress rather than division.
	
Section 2 formulates an extended CES production function. Section 3 explores wage dynamics and income distribution. Section 4 examines equilibrium effects of AGI capital accumulation. Section 5 assesses AGI’s impact on productivity. Section 6 discusses policy responses, followed by conclusions in Section 7.

\section{Substitution vs. Complementarity of AGI and Human Labor}

To analyze whether AGI labor complements or substitutes human labor, we use an extended CES production function incorporating human labor, AGI labor, traditional capital, and AGI capital given by
\begin{equation}
	Y = A \left[ \delta_K K^{\rho} + \delta_{AGI} K_{AGI}^{\rho} + \delta_h L_h^{\rho} + \delta_{L_{AGI}} L_{AGI}^{\rho} \right]^{\frac{1}{\rho}},
\end{equation}
where $A$ represents total factor productivity, $\delta_K, \delta_{AGI}, \delta_h, \delta_{L_{AGI}}$ are the share parameters for traditional capital, AGI capital, human labor, and AGI labor, respectively. The parameter $\rho$ governs the elasticity of substitution $\sigma$ between inputs, which determines how easily one input can be replaced by another:
\begin{equation}
	\sigma = \frac{1}{1 - \rho}.
\end{equation}
The relationship between AGI labor ($L_{AGI}$) and human labor ($L_h$) depends on the elasticity of substitution $\sigma$. If $\rho = 1$ ($\sigma \to \infty$), AGI labor and human labor are perfect substitutes, leading to full automation and the direct displacement of human workers. When $0 < \rho < 1$ ($1 < \sigma < \infty$), AGI labor remains highly substitutable but does not completely replace human labor. The degree of displacement depends on relative productivity and cost structures. If $\rho < 0$ ($0 < \sigma < 1$), AGI labor and human labor act as complementary inputs, where increasing AGI labor enhances human labor productivity instead of replacing it. In the Cobb-Douglas case ($\rho \to 0$, $\sigma = 1$) \cite{cobb1928,stiefenhofer2024}, both inputs exhibit unitary elasticity, meaning proportional increases in both AGI and human labor lead to proportional output gains without displacement. The marginal productivity of human labor and AGI labor are given by
\begin{align}
	MP_{L_h} &= \frac{\partial Y}{\partial L_h} = \delta_h A L_h^{\rho-1} \left( \delta_K K^{\rho} + \delta_{AGI} K_{AGI}^{\rho} + \delta_h L_h^{\rho} + \delta_{L_{AGI}} L_{AGI}^{\rho} \right)^{\frac{1}{\rho} - 1}, \\
	MP_{L_{AGI}} &= \frac{\partial Y}{\partial L_{AGI}} = \delta_{L_{AGI}} A L_{AGI}^{\rho-1} \left( \delta_K K^{\rho} + \delta_{AGI} K_{AGI}^{\rho} + \delta_h L_h^{\rho} + \delta_{L_{AGI}} L_{AGI}^{\rho} \right)^{\frac{1}{\rho} - 1}.
\end{align}
The relative marginal productivity determines whether AGI labor substitutes or complements human labor. The wage ratio is defined as
\begin{equation}
	\frac{w_{AGI}}{w_h} = \frac{MP_{L_{AGI}}}{MP_{L_h}},
\end{equation}
which implies that as AGI labor increases, its relative wage rises if it is a strong substitute for human labor. Conversely, in a complementary scenario, increased AGI labor enhances $MP_{L_h}$, stabilizing or even increasing human wages. The broader economic implications of AGI-human substitution patterns include wage dynamics, employment trends, and productivity growth. When $\sigma > 1$, human wages decline relative to AGI wages, leading to increased inequality unless redistributive measures are introduced. If AGI labor is highly substitutive ($\sigma \to \infty$), human employment shrinks, necessitating policy interventions such as Universal Basic Income ($\mathcal{U}$) or re-skilling programs. If AGI and human labor are complements ($\sigma < 1$), AGI adoption enhances productivity without displacing workers, fostering inclusive economic growth. Expanding AGI capital ($K_{AGI}$) shifts firms away from labor-intensive production methods, with the extent of this shift governed by $\rho$. Ultimately, the economic impact of AGI is critically dependent on the substitution elasticity $\sigma$. Policy interventions must align with the observed substitution patterns, ensuring that efficiency gains do not come at the cost of equitable labor market transitions.

\section{Wage and Income Distribution in an AGI Economy}
The marginal productivity of each factor determines income distribution in a competitive market. The marginal productivity of human labor is given by
\begin{equation}
	MP_{L_h} = A \delta_h L_h^{\rho} \left[ \sum \delta_i X_i^{\rho} \right]^{\frac{1}{\rho} - 1} \frac{1}{L_h}.
\end{equation}
Similarly, for AGI labor, we have
\begin{equation}
	MP_{L_{AGI}} = A \delta_{L_{AGI}} L_{AGI}^{\rho} \left[ \sum \delta_i X_i^{\rho} \right]^{\frac{1}{\rho} - 1} \frac{1}{L_{AGI}}.
\end{equation}
The marginal productivity of traditional capital is given by
\begin{equation}
	MP_K = A \delta_K K^{\rho} \left[ \sum \delta_i X_i^{\rho} \right]^{\frac{1}{\rho} - 1} \frac{1}{K},
\end{equation}
and the marginal productivity of AGI capital is
\begin{equation}
	MP_{K_{AGI}} = A \delta_{AGI} K_{AGI}^{\rho} \left[ \sum \delta_i X_i^{\rho} \right]^{\frac{1}{\rho} - 1} \frac{1}{K_{AGI}},
\end{equation}
where $X_i$ represents the set of all productive inputs contributing to output:
\begin{equation}
	X = \{ K, K_{AGI}, L_h, L_{AGI} \}.
\end{equation}
Thus, $X_i$ corresponds to traditional capital ($K$), AGI capital ($K_{AGI}$), human labor ($L_h$), and AGI labor ($L_{AGI}$), aggregating the economy's productive capacity within the CES framework. In the limit where AGI capital accumulates indefinitely ($K_{AGI} \to \infty$), the marginal productivity of human labor declines asymptotically:
\begin{equation}
	\lim_{K_{AGI} \to \infty} MP_{L_h} = 0.
\end{equation}
Since wages are determined by marginal productivity under competitive market conditions, this leads to a collapse in human wages
\begin{equation}
	\lim_{K_{AGI} \to \infty} w_h = 0.
\end{equation}
This result suggests a fundamental restructuring of income distribution, wherein AGI capital owners capture a growing share of economic output, displacing traditional labor as a primary source of income. The increasing inequality necessitates alternative mechanisms to sustain consumption and demand.

\section{Equilibrium Analysis and Long-Run Market Dynamics}
In the long-run equilibrium, as AGI capital ($K_{AGI}$) increases indefinitely, the marginal productivity of human labor ($MP_{L_h}$) asymptotically approaches zero
\begin{equation}
	\lim_{K_{AGI} \to \infty} MP_{L_h} = 0.
\end{equation}
Since wages are determined by marginal productivity under competitive market conditions, this leads to a collapse in human wages
\begin{equation}
	\lim_{K_{AGI} \to \infty} w_h = 0.
\end{equation}
This signifies a fundamental structural shift in the economy, where AGI capital supplants human labor as the primary driver of production. As labor-based income distribution becomes unsustainable, alternative mechanisms are required to sustain human consumption. These include
\begin{equation}
	C_h = \mathcal{T} + T_{AGI} + \mathcal{U},
\end{equation}
where $C_h$ represents human consumption, supported by wealth transfers ($\mathcal{T}$), progressive AGI capital taxation ($T_{AGI}$), and Universal Basic Income ($\mathcal{U}$). The marginal productivity of human labor is given by
\begin{equation}
	MP_{L_h} = A \delta_h L_h^{\rho - 1} \left( \sum \delta_i X_i^{\rho} \right)^{\frac{1}{\rho} - 1}.
\end{equation}
As $K_{AGI}$ increases, its contribution dominates the summation
\begin{equation}
	\sum \delta_i X_i^{\rho} \approx \delta_{AGI} K_{AGI}^{\rho} \quad \text{for large } K_{AGI}.
\end{equation}
Thus, the marginal productivity simplifies to
\begin{equation}
	MP_{L_h} \approx A \delta_h L_h^{\rho - 1} \left( \delta_{AGI} K_{AGI}^{\rho} \right)^{\frac{1}{\rho} - 1}.
\end{equation}
Taking the limit yields
\begin{equation}
	\lim_{K_{AGI} \to \infty} MP_{L_h} = 0.
\end{equation}
This confirms that as AGI capital expands, human labor's contribution to production diminishes, driving labor wages to zero. The economic implications of this transition are significant: wage collapse occurs as $MP_{L_h} \to 0$, shifting economic power toward AGI capital owners. Reduced wages lead to weakened aggregate demand, requiring redistribution mechanisms to maintain market stability. Additionally, the production function transforms, effectively eliminating human labor from the economic framework.
\begin{figure}[H]
	\centering
	\begin{minipage}{0.45\textwidth}
		\centering
		\includegraphics[width=\textwidth]{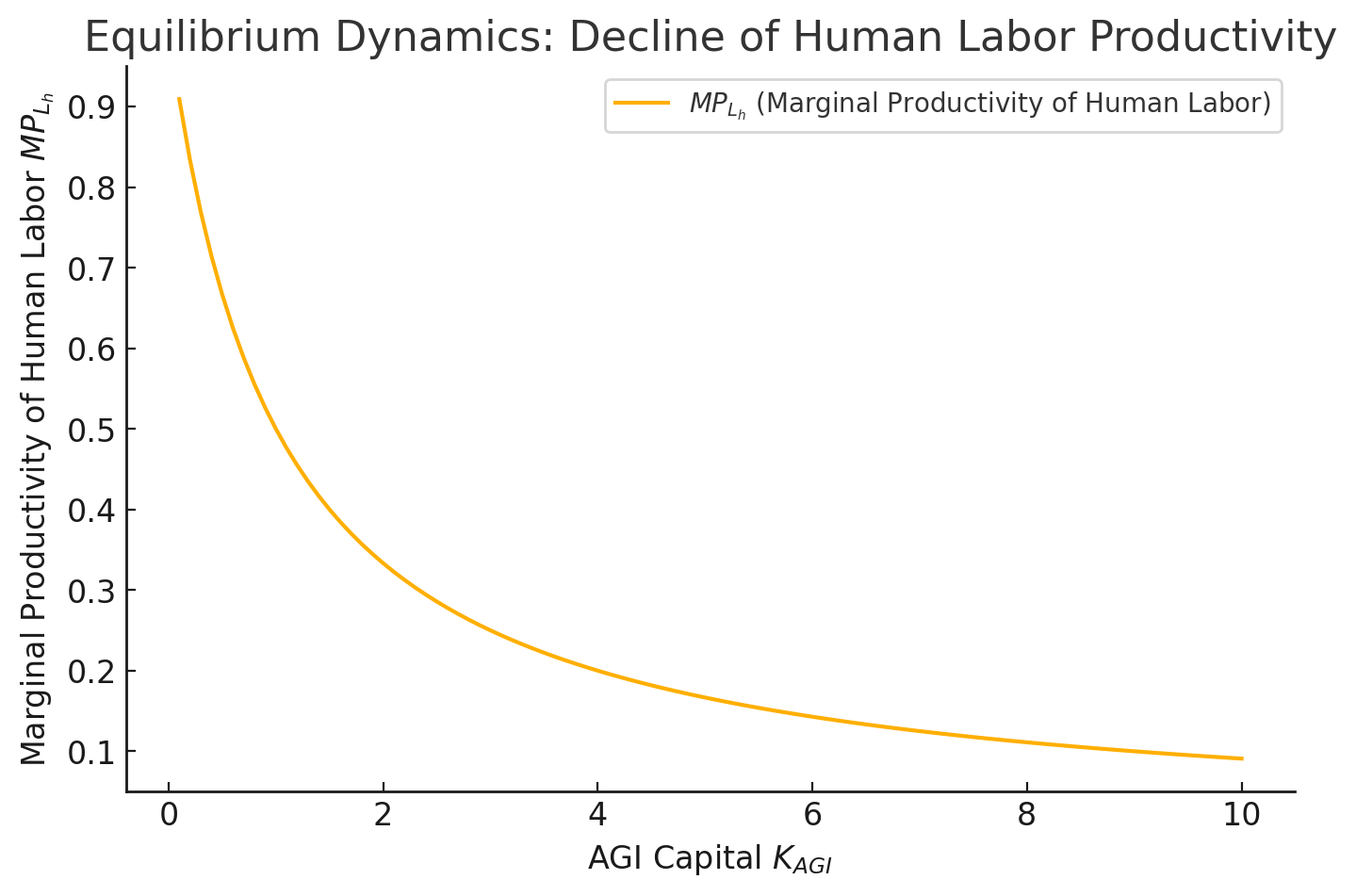}
	\end{minipage}
	\hfill
	\begin{minipage}{0.45\textwidth}
		\centering
		\includegraphics[width=\textwidth]{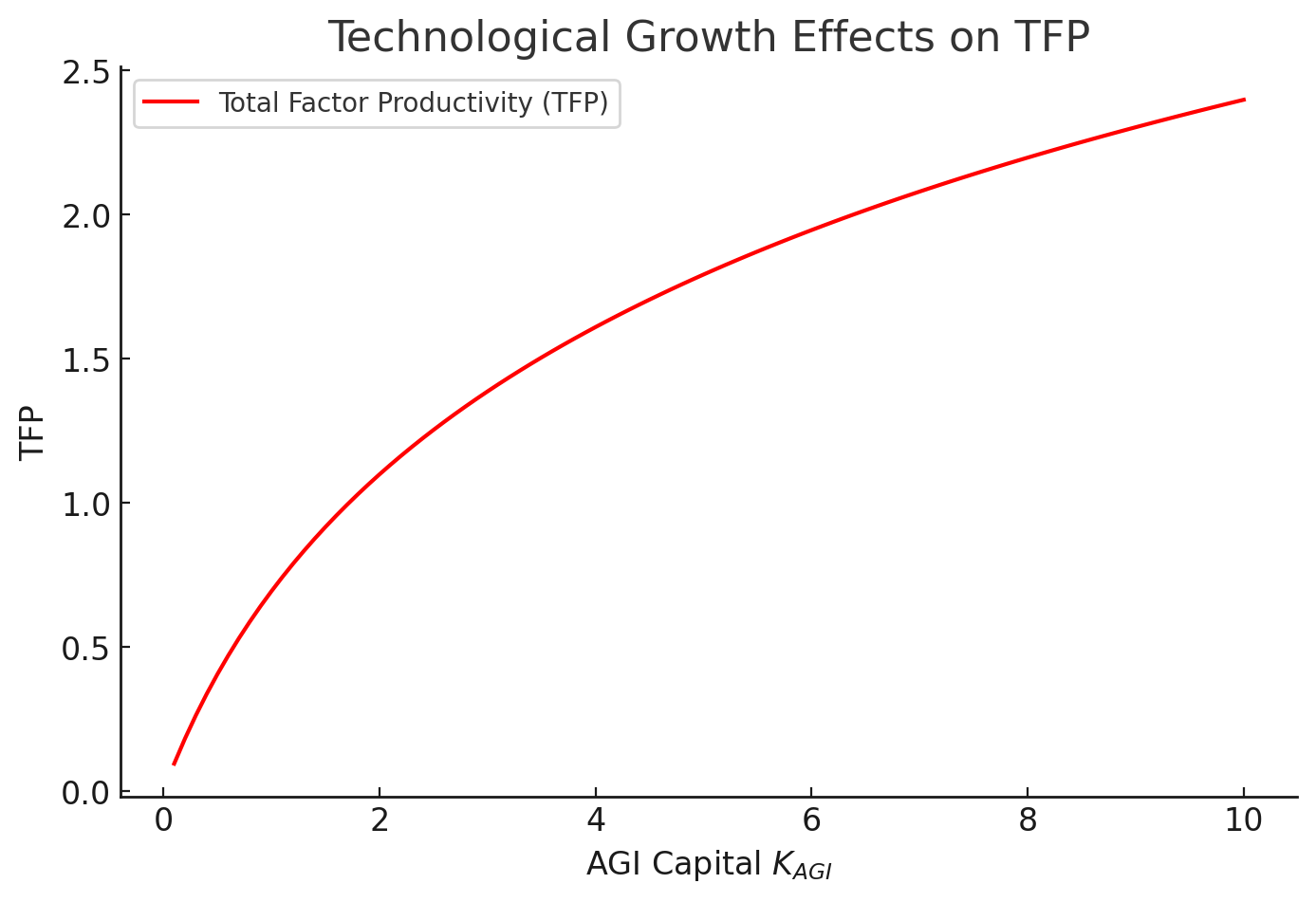}
	\end{minipage}
	\caption{(Left)The decline in human labor productivity as AGI capital $K_{AGI}$ increases, illustrating the potential collapse of human wages and the need for alternative economic structures. (Right) The impact of AGI capital on total factor productivity (TFP), where growth depends on whether AGI complements or substitutes human labor.}
\end{figure}

\section{Technological Growth and Productivity Effects}
Total Factor Productivity (TFP) serves as a fundamental measure of technological efficiency, capturing the residual contribution of capital and labor inputs to economic output. It is formally defined as
\begin{equation}\label{TFP}
	TFP = \frac{Y}{K + K_{AGI} + L_{AGI} + L_h}.
\end{equation}
Substituting the CES)production function into (\ref{TFP}) yields
\begin{equation}
	TFP = A \frac{\left( \sum_{i} \delta_i X_i^{\rho} \right)^{\frac{1}{\rho}}}{K + K_{AGI} + L_{AGI} + L_h}.
\end{equation}
The asymptotic behavior of TFP depends on the elasticity of substitution $\sigma = \frac{1}{1-\rho}$. If $\sigma < 1$ (i.e., strong complementarity), AGI-driven inputs amplify human productivity, yielding sustained technological growth as $K_{AGI} \to \infty$
\begin{equation}
	\lim_{K_{AGI} \to \infty} TFP = A \left( \delta_h L_h^{\rho} + \delta_K K^{\rho} + \delta_{AGI} K_{AGI}^{\rho} \right)^{\frac{1}{\rho}}.
\end{equation}
Conversely, if AGI capital substitutes human labor ($\sigma > 1$), increasing $K_{AGI}$ reduces human labor’s marginal contribution, causing wage suppression and stagnating aggregate productivity due to demand contraction. In this case
\begin{equation}
	\lim_{K_{AGI} \to \infty} \frac{MP_{L_h}}{MP_{K_{AGI}}} = 0, \quad \text{leading to } \quad \lim_{K_{AGI} \to \infty} w_h = 0.
\end{equation}
A wealth concentration threshold emerges where AGI capital owners disproportionately capture income, reducing overall economic efficiency. This bifurcation manifests when labor income as a share of total output vanishes
\begin{equation}
	\lim_{K_{AGI} \to \infty} \frac{w_h L_h}{Y} = 0.
\end{equation}
Sustainable growth requires redistributive policies, ensuring consumption stability despite labor market disruption. A macroeconomic equilibrium condition ensuring nonzero consumption for human workers is given by
\begin{equation}
	\int_0^\infty C_h(t) dt = \int_0^\infty \left(\mathcal{T}(t) + T_{AGI}(t) + \mathcal{U}(t)\right) dt,
\end{equation}
where $C_h$ is human consumption, supported by targeted redistribution mechanisms ($\mathcal{T}$), taxation ($T_{AGI}$), and universal basic income ($\mathcal{U}$). This refined analysis demonstrates that AGI capital expansion induces two possible economic trajectories: (1) sustained TFP growth if AGI complements human labor or (2) stagnation and inequality if substitution dominates. In the latter case, redistributive interventions such as Universal Basic Income ($\mathcal{U}$) or progressive taxation ($T_{AGI}$) become critical for preserving economic stability and aggregate demand. Without these mechanisms, excessive capital concentration leads to bifurcated economic gains, disproportionately benefiting AGI capital owners while eroding labor income and market equilibrium.
\begin{figure}[H]
	\centering
	\begin{minipage}{0.45\textwidth}
		\centering
		\includegraphics[width=\textwidth]{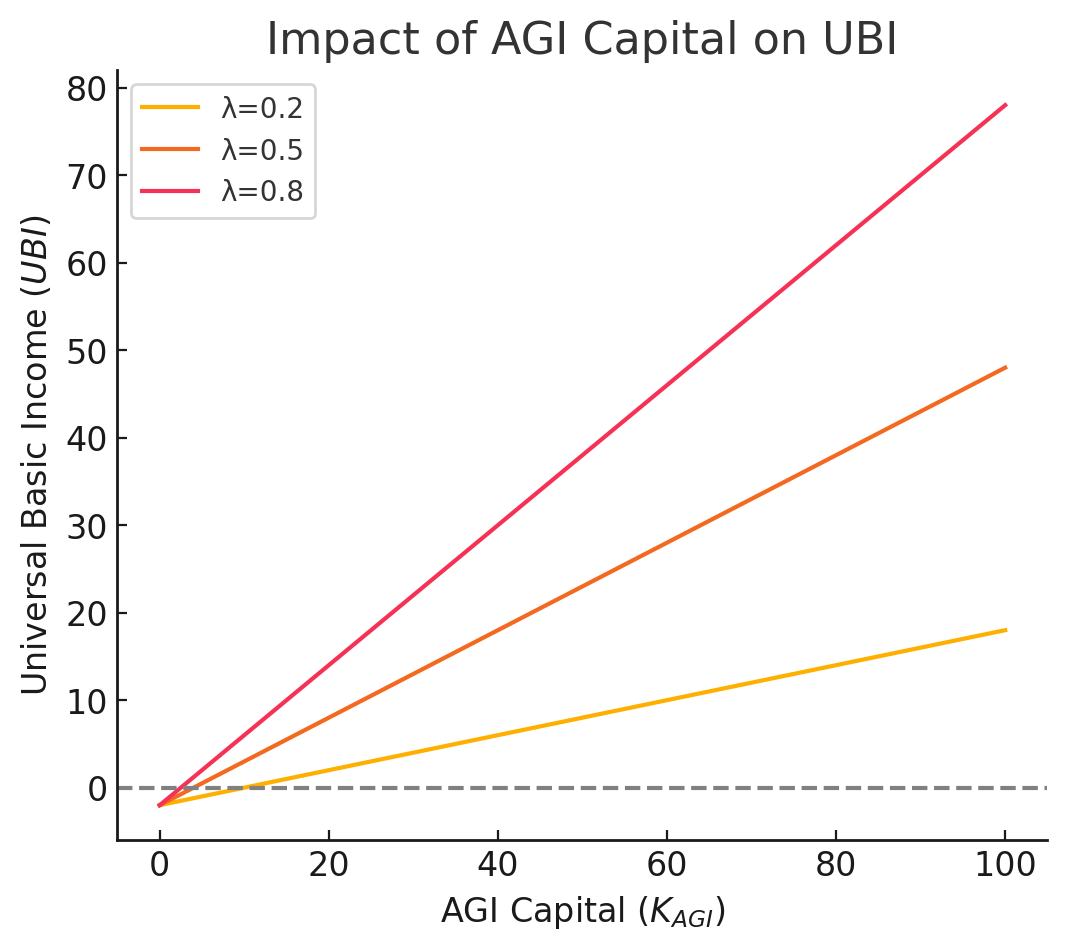}
	\end{minipage}
	\hfill
	\begin{minipage}{0.45\textwidth}
		\centering
		\includegraphics[width=\textwidth]{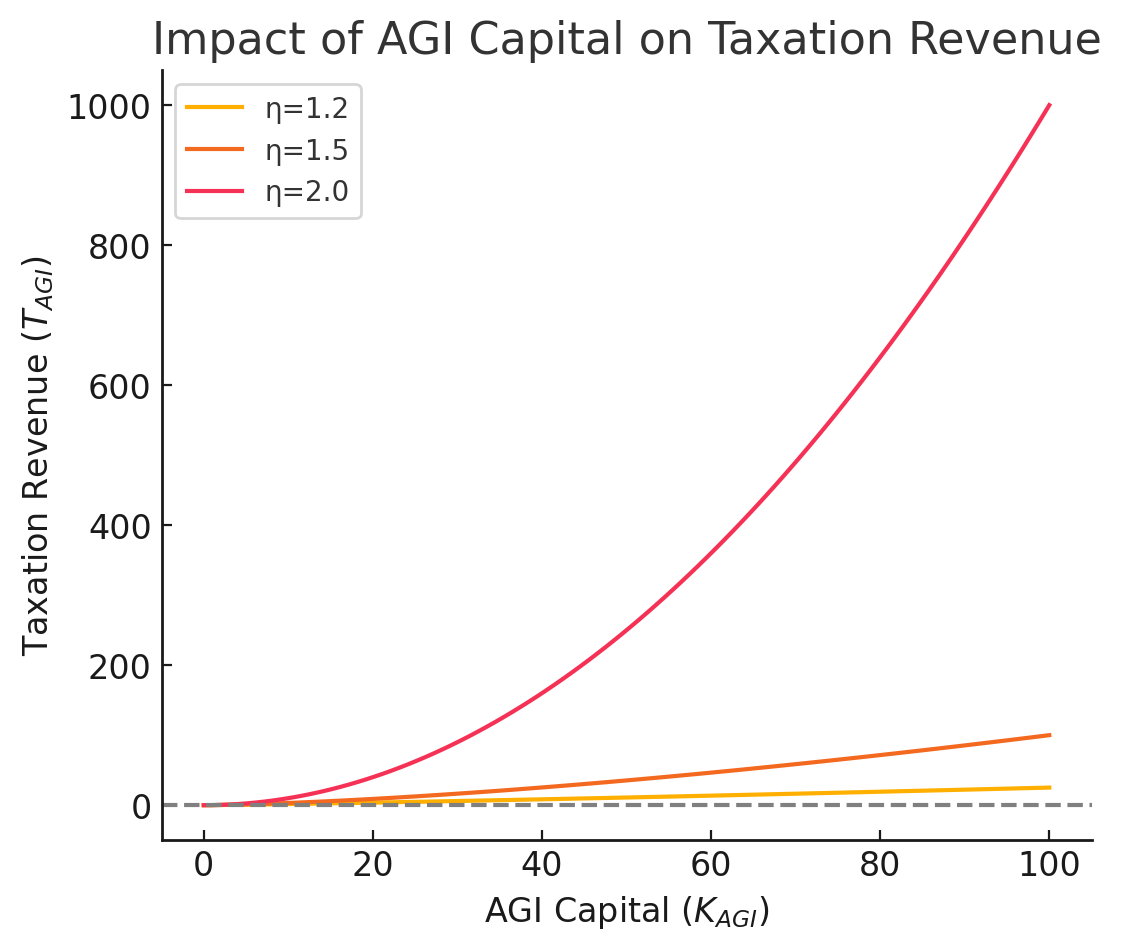}
	\end{minipage}
	\caption{(Left) This plot illustrates how Universal Basic Income (UBI) changes with increasing AGI capital ($K_{AGI}$) under different redistribution fractions ($\lambda$). Higher $\lambda$ values result in greater UBI, ensuring a more substantial redistribution of AGI-generated output. However, administrative costs ($C_{UBI}$) slightly offset the total amount available for distribution.	(Right) This plot shows the relationship between AGI capital ($K_{AGI}$) and taxation revenue ($T_{AGI}$) under different levels of tax progressivity ($\eta$). As $\eta$ increases, taxation on AGI capital grows non-linearly, enhancing redistribution while maintaining investment incentives. The results highlight how progressive taxation can be used to manage wealth concentration in an AGI-driven economy. }
\end{figure}

\section{Government Policy and Economic Interventions}
To mitigate economic instability induced by AGI-driven labor displacement, policy interventions must ensure equitable redistribution of productivity gains while preserving long-term economic efficiency. A key mechanism is Universal Basic Income (UBI), which can be structured as a function of AGI-driven output:
\begin{equation}
	UBI = \lambda Y_{AGI} - C_{UBI},
\end{equation}
where $\lambda \in (0,1)$ represents the fraction of AGI-generated output redistributed as basic income, and $C_{UBI}$ denotes administrative and operational costs. The sustainability of UBI depends on the optimal choice of $\lambda$ that maximizes social welfare while maintaining incentives for innovation. A complementary approach is the implementation of progressive AGI capital taxation, ensuring equitable wealth distribution while preserving investment incentives. The tax function is given by
\begin{equation}
	T_{AGI} = \tau K_{AGI}^{\eta},
\end{equation}
where $\tau$ is the baseline tax rate and $\eta > 1$ controls tax progressivity. Higher values of $\eta$ impose steeper taxation on AGI capital, promoting redistribution while avoiding excessive capital accumulation deterrence. Public or cooperative ownership of AGI capital can serve as an alternative mechanism to balance economic efficiency and equity. The optimal allocation of AGI capital should maximize social welfare, which can be formalized as a constrained welfare maximization problem is to
\begin{equation}
	\max_{K_{AGI}^{pub}} W = U(Y) - D(K_{AGI}^{pub}),
\end{equation}
subject to
\begin{equation}
	Y = A \left( \delta_K K^{\rho} + \delta_{AGI} K_{AGI}^{\rho} + \delta_h L_h^{\rho} + \delta_{L_{AGI}} L_{AGI}^{\rho} \right)^{\frac{1}{\rho}}.
\end{equation}
Here, $W$ represents aggregate social welfare, $U(Y)$ denotes utility derived from economic output, and $D(K_{AGI}^{pub})$ captures the costs associated with public deployment of AGI capital. The optimal public ownership level $K_{AGI}^{pub*}$ satisfies the first-order condition
\begin{equation}
	\frac{dW}{dK_{AGI}^{pub}} = \frac{dU}{dY} \cdot \frac{dY}{dK_{AGI}^{pub}} - \frac{dD}{dK_{AGI}^{pub}} = 0.
\end{equation}
This ensures that the marginal benefit of increasing public AGI capital equals its marginal cost. Efficient allocation requires designing $D(K_{AGI}^{pub})$ such that excessive state control does not stifle innovation while ensuring a socially optimal balance of AGI capital ownership. This analysis underscores the necessity of redistributive interventions to prevent economic bifurcation. Policies such as Universal Basic Income ($\mathcal{U}$), progressive taxation ($T_{AGI}$), and public AGI capital ownership ensure that economic gains are equitably distributed while sustaining aggregate demand and preventing excessive wealth concentration among AGI capital owners.

\section{Conclusion}
This study provides a theoretical foundation for understanding how AGI-driven labor and capital transform economic production, labor markets, and technological growth. By extending the CES production function to incorporate AGI labor and AGI capital, we capture the dynamic interplay between automation, human employment, and long-term productivity. 

Our analysis highlights several key insights. First, the degree of substitution between AGI and human labor plays a decisive role in determining economic outcomes. If AGI labor is a strong substitute for human labor, wages decline, and traditional employment structures become unsustainable. Conversely, if AGI labor complements human labor, productivity gains can be widely shared, leading to enhanced economic growth without severe displacement\cite{acemoglu2022}. Second, as AGI capital accumulates, the marginal productivity of human labor may approach zero, necessitating alternative income distribution mechanisms such as Universal Basic Income (UBI) or progressive AGI taxation to sustain aggregate demand\cite{loebbing2022,thuemmel2023}. Third, the long-run stability of the economy depends on whether AGI capital enhances total factor productivity or leads to stagnation due to weakened labor-driven consumption \cite{acemoglu2011,acemoglu2019}.

This research underscores the profound and far-reaching implications of AGI for society. As AGI transforms the foundations of labor, capital, and productivity, the traditional economic structures that have long governed human progress may no longer suffice. The emergence of AGI challenges our fundamental assumptions about work, value, and wealth distribution, raising existential questions about the role of human labor in an increasingly automated world.

In light of these disruptions, a new Social Contract is not just desirable—it is essential. The balance between technological efficiency and social equity must be redefined to ensure that AGI serves as a catalyst for shared prosperity rather than a force of deepening economic stratification. Governments and institutions must adopt forward-thinking regulatory frameworks that accommodate the realities of automation while safeguarding human dignity and agency.

More than just an economic adjustment, this shift requires a philosophical reexamination of how societies allocate wealth, opportunity, and purpose. Future research must move beyond theoretical models to explore how real-world economies adapt to AGI’s transformative potential. The question is no longer whether AGI will disrupt society, but rather how societies will choose to respond—whether they will passively endure its consequences or actively shape a future where technology aligns with human flourishing.

\end{document}